# Breakdown of the integer and fractional quantum Hall states in a quantum point contact

C. Dillard,[1] X. Lin,[1,2,*] M. A. Kastner,[1] L. N. Pfeiffer,[3] and K. W. West[3]

[1]*Department of Physics, Massachusetts Institute of Technology, Cambridge, Massachusetts 02139, USA*

[2]*International Center for Quantum Materials, Peking University, Beijing, People's Republic of China 100871*

[3]*Department of Electrical Engineering, Princeton University, Princeton, New Jersey 08544, USA*

\* Corresponding author, linxi07@gmail.com

Abstract:

The integer and fractional quantum Hall states are known to break down at high dc bias, exhibiting deviation from the ideal incompressible behavior. We measure breakdown of the ν = 2, 3, 4, 5 integer and the ν = 4/3 and 5/3 fractional states in a quantum point contact (QPC) of lithographic width ~600 nm. Dependence of the critical current on magnetic field, QPC gate voltage, and QPC width are presented. Of particular interest, the critical current of the 4/3 and 5/3 fractional states shows the opposite dependence on QPC width compared to the integer states. This previously unobserved result is not explained by current theories of breakdown.

## I. INTRODUCTION

A two-dimensional electron system (2DES) under a perpendicular magnetic field exhibits the integer quantum Hall effect (QHE) at low temperatures,[1] when the thermal excitation becomes much smaller than the Landau level energy gap. In such a system, when the Fermi energy lies between Landau levels the longitudinal resistance $R_{xx}$ vanishes while the transverse, or Hall, resistance becomes quantized at $R_{xy} =(1/f)(h/e^2)$, where $f$ is equal to the number of filled Landau levels. The state of the system is usually characterized by the filling factor $v = n/(B/\Phi_0)$, where $n$ is the electron density, $B$ is the magnetic field, and $\Phi_0 = h/e$ is the quantum of flux. The fractional QHE[2] arises when electrons form composite fermions by binding to flux quanta.[3] These composite fermions in turn exhibit their own version of the QHE at fractional filling factors with odd denominators. Similar to the integer QHE, $R_{xy}$ forms plateaus at values $R_{xy} = (1/f)(h/e^2)$ (in this case $f$ is a fraction) with $R_{xx} = 0$. However, unlike the integer QHE, the fractional version is fundamentally a result of collective interactions.

In both the integer and fractional regimes, quantized quantum Hall states will not be observed if electrons or quasi-particles are able to access extended states other than the ground state. This can happen, for example, if the temperature becomes comparable to the energy gap of the system.[4] Alternatively, a dc bias will cause breakdown of the quantum Hall states if the current through the sample exceeds some critical value.[5,6] Multiple studies of this type of breakdown have been performed, mostly focused on the integer QHE regime; see the review by Nachtwei for examples of earlier work.[7] The value of the critical current is sample-dependent and has been found to have either a linear[8–15] or sublinear[12–16] dependence on sample width. Breakdown has been observed in the fractional QHE regime,[17,18] but to our knowledge no dependence of the critical current on sample width has been reported. A number of explanations for the breakdown have been proposed, but no consensus seems to have been reached.[7] Understanding the mechanism by which quantum Hall states break down is important not only to achieve a better understanding of the fundamental physics of the QHE but also to ensure that measurements of the QHE as the standard of resistance can be made as accurately as possible.

## II. EXPERIMENTAL DETAILS

We present measurements of a quantum point contact (QPC) formed in a high-mobility MBE-grown GaAs/AlGaAs heterostructure. At base temperature the sample mobility is measured to be $1 \times 10^7$ cm$^2$ V$^{-1}$ s$^{-1}$ and the electron density to be $2.6 \times 10^{11}$ cm$^{-2}$. A 30 nm wide quantum well lies 200 nm below the surface with Si δ-donors within thin GaAs layers residing at 100 nm above and below the center of the quantum well. A 150 µm wide Hall bar is formed by etching, and in the middle of this a QPC is created by applying a voltage to Cr/Au top gates. Applying a negative voltage to these gates depletes the electrons from the 2DES underneath, creating a constriction. A false-color image of the active region of the device is shown in Fig. 1(a). The two light grey gates are energized to form the QPC, while the remaining four dark grey gates are kept grounded during the experiment. The lithographic width of the QPC is approximately 600 nm. The geometry of this constriction is similar to those used by Bliek et al.[19] and Kawaguchi et al.;[11] however, we have an added advantage of being able to change the width of the constriction by varying the voltage applied to the gates. The sample is cooled in a dilution refrigerator with a base electron temperature of ~13 mK.

A schematic of the mesa and measurement setup is shown in Fig. 1(b). An ac current of 0.4 nA RMS is sourced on one end of the Hall bar and the other end is grounded. Measurements of differential resistances $r_{xx}$, $r_{xy}$, and $r_d$ (as shown) are performed using standard lock-in techniques at 17 Hz. The local physics of the QPC are probed by $r_d$, which is measured across the Hall bar and on opposite sides of the QPC. Conversely, $r_{xx}$ and $r_{xy}$ are insensitive to the QPC and, instead, depend on the large-scale physics of the full Hall bar. Breakdown is studied by applying a dc current $I_{dc}$ of up to ±100 nA. Previous studies of breakdown of quantum Hall states have mostly focused on measurements of the longitudinal resistivity $\rho_{xx}$, or similar quantities, measured using voltage contacts on the same side of the Hall bar but on either side of the region of breakdown. This differs from our $r_{xx}$, which is measured far away from the QPC and reflects the behavior of the Hall bar as a whole. However, our measurements are performed in regimes in which $r_{xy}$ is well-quantized and constant. Hence the quantity $r_d - r_{xy}$ should exhibit similar behavior to the $\rho_{xx}$ measured by others.

We observe breakdown of the quantum Hall states in the QPC at dc biases for which $r_{xx}$ and $r_{xy}$ show no sign of breakdown outside the QPC. However, in order for this comparison to be meaningful, the electron density in the region inside the QPC needs to be equal to the density far away in the Hall bar. Because the magnetic field is constant throughout the sample, a uniform electron density will produce a uniform filling factor, allowing for valid comparison between $r_d$ and $r_{xy}$. Normally, energizing the QPC gates reduces the electron density in the QPC region to less than that in the surrounding Hall bar. For example, applying gate voltages similar to those used in this study in the usual way at base temperature would reduce the electron density from 2.6 to $2.2 \times 10^{11}$ cm$^{-2}$, more than a 15% reduction. We avoid this problem by using the technique of annealing the gates at 4 K.[20] A gate voltage of -2.7 V is applied at 4 K for ~40 h before cooling to base temperature. After cooling, the gate voltage $V_G$ is limited to between -2.7 and -2.0 V. Fig. 2 demonstrates that this annealing technique is effective in maintaining a uniform electron density throughout the sample. The electron density in the QPC and far away from the QPC can be compared using the slopes of $r_d$ and $r_{xy}$, respectively, at low magnetic field. In addition, when the density is uniform the features of the Hall trace, such as the endpoints of the plateaus, occur at equal magnetic fields for $r_d$ and $r_{xy}$. Between the plateaus, $r_d$ exhibits increased resistance, because of backscattering in the QPC. At $V_G$ = -2.7 V the density in the QPC is slightly less than that elsewhere in the Hall bar, but significantly closer than the 15% difference found without annealing. Even this difference goes away for gate voltages more positive than -2.4 V. Measurements of breakdown are performed with -2.4 V $\leq V_G \leq$ -2.0 V. In summary, annealing creates an electron density in the QPC within 1% of the density throughout the Hall bar for the gate voltages used in our measurements. We suspect that the annealing results from redistribution of electrons among the donors, leading to a constant electrochemical potential and uniform maximum density.

## III. RESULTS

The breakdown feature we observe is shown in Fig. 3(a). Differential resistances $r_d$, $r_{xy}$, and $r_{xx}$ versus $I_{dc}$ are measured simultaneously at various magnetic fields on the high-field side of the ν = 2 plateau. On the plateau (black curves), $r_{xy}$ is well-quantized and $r_{xx} = 0$ within the noise. As $B$ is increased past the plateau (grey curves) both $r_{xy}$ and $r_{xx}$ show deviations from this behavior. In contrast, $r_d$ increases from the quantized value at lower magnetic fields, showing an abrupt rise, with, in many cases, a sharp peak, and a high-dc-bias value larger than the quantized plateau resistance. Because we are measuring the differential resistance, the sharp peak corresponds to a steep rise in the current above threshold. The critical current $I_c$, at which the sharp rise in differential resistance occurs, is generally well-defined and we use the value of $I_{dc}$ at which $r_d$ rises above the noise on the plateau. Figure 3(b) shows similar measurements of $r_d$ over the entire $\nu = 5/3$ plateau. Here the breakdown does not exhibit a peak and the rise off of the quantized plateau is not as sharp. However, we can still accurately define the critical current as the point at which $r_d$ exceeds the noise on the plateau.

We measure $I_c$ as a function of magnetic field within each plateau and of gate voltage at fixed magnetic field. In each case we sweep $I_{dc}$ between ±100 nA, alternating directions, while stepping $B$ or $V_G$ after each sweep. For each value of $B$ or $V_G$ we calculate $|I_c|$ as described above; the spread in $|I_c|$ at each point is predominantly the result of hysteresis in the $I_{dc}$ sweep direction. We observe two different types of hysteresis, each of which seems to occur inconsistently. One is similar to that observed in previous experiments,[6,14,21] illustrated in Fig. 4(a). This type of hysteresis may be explained by the electron heating model of breakdown;[22] however, as discussed below, the electron heating model cannot explain our results in the fractional quantum Hall regime. We also observe a strange type of hysteresis illustrated in Fig. 4(b). In some cases, both types of hysteresis occur simultaneously. For all measurements, the $I_{dc}$ sweep rate is 0.7 nA/s. We have tested the dependence on sweep rate at filling factors 2 and 5, sweeping $I_{dc}$ with fixed $B$ and $V_G$, and find no significant differences for slower sweep rates down to $\sim 1 \times 10^{-3}$ nA/s.

Critical currents for integer and fractional filling factors are shown in Fig. 5 as a function of $B$ and for two values of $V_G$. For integer filling factors, $I_c$ is strongly nonlinear with magnetic field. Similar nonlinearities have been observed in GaAs samples of width $\sim 1$ μm[19,23] and 380 μm[6] at $\nu = 2$, and in graphene,[24] although other studies have observed different (mostly linear) behavior.[25–28] The dependence on magnetic field exhibits power law behavior, at least over a limited range. Least-squares fits are performed using a power law equation

$$|I_c| = A\ (B_0 - B)^\alpha \tag{1}$$

with $B_0$ fixed to the end of the plateau. The resulting best-fit values of the power α are listed in Table I. More details about the fits are included in the Supplemental Material. As $B$ is decreased towards the center of each plateau, $|I_c|$ increases beyond the 100 nA range over which we measure. Although we are still able to measure $|I_c|$ near to the center of the plateau, the strong nonlinearity of $I_c$ makes it hard to estimate the maximum value of $|I_c|$, which presumably occurs at the center of each plateau. We do also observe deviations of $r_d$ from the quantized value on the low-field side of integer plateaus. However, the breakdown features are qualitatively different on the low-field side of the integer plateaus, as illustrated in Fig. 6. On the low-field side $|I_c|$ increases even more rapidly with increasing $B$, exceeding 100 nA within 20% of the plateau width. In addition, we find that $r_{xy}$ and $r_{xx}$ also show deviations at nearly the same values of $I_{dc}$. It appears that on the low-field side of integer QHE plateaus, the QPC and the full Hall bar undergo similar transitions out of the incompressible state. This is in contrast to the high field side of the plateaus where $|I_c|$ is clearly lower for $r_d$ than for $r_{xy}$. Hence we do not plot $|I_c|$ values for the low field side of the integer plateaus in Fig. 5. Finally, we note that on the low-field side of the integer plateaus, $r_d$ decreases rather than increases for $|I_{dc}| > I_c$.

We observe breakdown at fractional filling factors of 4/3 and 5/3, but not at 5/2, which instead exhibits a zero-bias peak due to backscattering at the QPC.[20] The behavior of breakdown at fractional filling factors 4/3 and 5/3 is different from that at the integers in three ways. First, $|I_c| \leq 100$ nA over the entire plateau, allowing us to measure the maximum $|I_c|$ near the middle of the plateau. Second, $r_d$ qualitatively maintains the same behavior ($r_d$ increases for $|I_{dc}| > |I_c|$) along the entire plateau. This is in contrast to the integers, which show qualitatively different behavior on the high and low field sides, as described above. Last, $|I_c|$ exhibits the opposite dependence on $V_G$, as indicated by the black ($V_G = -2.4$ V) and grey ($V_G = -2.2$ V) curves. For integer filling factors, more positive gate voltage leads to greater $|I_c|$, while for the fractions $|I_c|$ decreases or remains the same.

We can compare the maximum critical currents in the fractional plateaus to values measured by others. Takamasu et al. find a relationship between the critical currents at various integer and fractional filling factors:

$$I_c \propto |B_{eff}/B|, \qquad (2)$$

with $B_{eff}$ the effective magnetic field.[17] For integer states $B_{eff} = B$. However, the composite fermions which form the fractional QHE states experience a reduced magnetic field;[29] for the $\nu = 4/3$ and $5/3$ states $B_{eff} = |B - n\Phi_0/(3/2)|$. Equation (2) implies that the $4/3$ and $5/3$ states would exhibit the same critical current. However, we find that $|I_c|$ is significantly larger for $\nu = 5/3$, in contradiction to the results of Takamasu et al.[17]

Figure 7 more directly exhibits the dependence of $|I_c|$ on gate voltage at fixed magnetic field for each filling factor. Each of the four integer states exhibit increasing $|I_c|$ with more positive $V_G$, while the fractional states have the opposite dependence. Making the gate voltage more negative has two possible consequences: changing the electron density in the QPC and changing the channel width. We are able to estimate the channel width and electron density as a function of $V_G$ using the behavior of $r_d$ near zero magnetic field. Unlike $r_{xy}$, $r_d$ is nonzero at zero field; the increased resistance is caused by the QPC constriction. The combination of the electrostatic confinement and magnetic field leads to the formation of magnetoelectric subbands which are depopulated with increasing magnetic field.[30] For different potential profiles, one can calculate the resistance through an infinitely long channel as a function of $B$.[30,31] We fit two different models to the low field $r_d$ data to extract the QPC width and the electron density in the QPC. We assume either a square well potential or a parabolic potential. Results of the two approaches are shown in Fig. 8, along with the electron density in the Hall bar calculated from $r_{xy}$. Although the electron density in the QPC changes slightly with $V_G$, the change is not monotonic and the electron density in the QPC remains within 1% of that of the Hall bar over the range of gate voltages for which we measure $I_c$. The dependence of $I_c$ on $V_G$ observed for the fractional filling factors cannot be explained by the measured changes in electron density. For fixed filling factor, a change in electron density would produce the same effect as a change in magnetic field. However, because the electron density changes nonmonotonically with $V_G$, such changes cannot explain the monotonic change in $|I_c|$. In addition, even in regions with qualitative agreement, the magnitude of the change in electron density falls short by a factor of ~2 to 10 relative to that needed explain the change in $I_c$.

In contrast, the QPC width shows a strong dependence on $V_G$. At the annealing voltage of -2.7 V the fits give widths in rough agreement with the ~600 nm lithographic width of the QPC. Presumably, the annealing causes the channel width to be greater than it would otherwise be. As $V_G$ becomes more positive, the width increases to be on the order of a few micrometers. The widths extracted with the parabolic potential model are ~30% larger than those obtained using the square well model. This is consistent with the ratio of the fit functions in the $B = 0$ limit: the expression for the parabolic potential is larger by a factor of $4/\pi$. For both models the width increases linearly with more positive gate voltage. We perform a linear fit of QPC width versus $V_G$ and use the results to convert the $I_c$ dependence on $V_G$ of Fig. 7 to a dependence on QPC width. The results are shown in Fig. 9, using the width given by the square well model. The $\nu = 3, 4, 5$ states exhibit a linear dependence of $|I_c|$ on the QPC width, while $\nu = 2$ appears sublinear. Extrapolating to $|I_c| = 0$ for integer filling factors gives a nonzero intercept on the width axis on the order of 1 µm. Perhaps nonintuitively, we still observe a well-quantized QHE when the QPC width is less than 1 µm ($V_G = -2.7$ V), meaning that $|I_c| > 0$ even for widths less than those given by the intercepts in Fig. 9. The value of the intercept is similar for the $\nu = 3, 4$, and 5 states (for a given estimate of width), also consistent with previous experiment.[8,10] The fractional states show opposite behavior: $|I_c|$ decreases with increasing QPC width. This is the first reported result of width dependence of breakdown in the fractional QHE regime of which we are aware, and is in direct contrast to all previously reported results in the integer regime.

During a different cool-down of the same device we annealed at -3.0 V and measured temperature dependence of breakdown of the $\nu = 2$ state. Similar to the data displayed above, we measured $|I_c|$ versus $B$ at $V_G = -2.7$ V and $|I_c|$ versus $V_G$ at 5.8 T. Neither showed any temperature dependence between 20 and 80 mK. See the Supplemental Material for more details on our temperature dependence measurements.

IV. DISCUSSION

The dependence of critical current on QPC width observed for the fractional quantum Hall states (Fig. 9) cannot be explained by any theory of breakdown of which we are aware. Simply put, every theory predicts an increasing critical current with increasing width, while in the fractional regime we observe the opposite dependence. We briefly discuss a number of these theories and their predictions for the width dependence of the critical current.

Eaves and Sheard propose that breakdown is caused by quasi-elastic inter-Landau-level scattering.[32] In this process, an electron from the uppermost filled Landau level scatters into an empty state of equal energy in the next, unfilled, Landau level. Normally such scattering is highly suppressed by the energy gap between Landau levels; however if the Hall electric field is strong enough, two such states of equal energy may have enough overlap in their wave functions to allow scattering to occur. In this model, breakdown occurs at a critical Hall field, and hence the critical current increases with sample width. Shizuya considers what he refers to as an "intrasubband process" in which an increased Hall field causes delocalization of previously-localized impurity states[33]. The newly extended states carry current through the sample, leading to breakdown. Again, because of the role of the Hall field, the critical current is expected to increase with sample width. A number of authors have considered electron heating as a cause of breakdown.[6,22,34,35] Energy is gained from the electric field, with a rate $P = \mathbf{J} \cdot \mathbf{E} = \sigma_{xx}(E_x^2 + E_y^2) \approx \sigma_{xx}E_y^2$, and dissipated by some generally unspecified mechanism. Here, $\mathbf{J}$ is the current density, $\mathbf{E}$ is the electric field, and $\sigma_{xx}$ is the longitudinal conductivity. Reasonable assumptions for the dependence of the dissipation on the electron and lattice temperatures lead to runaway heating at a critical Hall field, which in turn causes breakdown.

Středa and von Klitzing consider an electron-phonon interaction, in which an electron moves from one edge of the sample to the other, with the excess energy absorbed by phonons.[36] This is allowed only when the drift velocity $v = E_y/B$ of electrons exceeds the sound velocity in GaAs. As with other theories, this implies the existence of a critical Hall field. Similarly, Martin et al. explore the formation of magnetoexcitons at a critical velocity or, equivalently, a critical Hall field.[37] Finally, Tsemekhman et al. consider a percolation model and calculate the critical Hall field at which a metallic path opens across the sample.[38] This metallic path is caused by an avalanche-type breakdown of successive incompressible regions, leading to breakdown of the sample as a whole. In all these cases the dependence of breakdown on a critical Hall field implies that the critical current should increase with sample width.

## V. CONCLUSION

We have studied breakdown of integer and fractional quantum Hall states induced by high dc current bias through a QPC formed in a high mobility GaAs heterostructure. Dependence of the critical current on magnetic field and gate voltage was measured for filling factors 4/3, 5/3, 2, 3, 4, and 5. We extract the QPC width as a function of gate voltage from low magnetic field measurements by fitting to various theoretical forms. Filling factors 3, 4, and 5 exhibit a linear dependence of critical current on the QPC width while $\nu = 2$ appears to be sublinear. In stark contrast the fractional filling factors exhibit a decreasing critical current with increasing QPC width. This behavior is the first reported width dependence of breakdown of fraction quantum Hall states of which we are aware, and is not explained by current theories of QHE breakdown.

We are grateful to Claudio Chamon, Dmitri Feldman, and Xiao-Gang Wen for helpful discussions. We thank Jeff Miller and Charles Marcus for sample fabrication, which was done at Harvard's Center for Nanoscale Systems with support from Microsoft Project Q. The work at MIT was funded by National Science Foundation under Grant No. DMR-1104394. The work at Princeton was partially funded by the Gordon and Betty Moore Foundation as well as the National Science Foundation MRSEC Program through the Princeton Center for Complex Materials (Grant No. DMR-0819860).

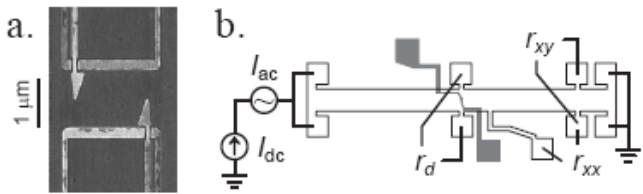

FIG. 1. Device image and measurement setup.

(a) False color scanning electron micrograph of a device fabricated similarly to the one used in this experiment. The light grey gates are annealed and biased negatively (see text for details) and the dark grey gates remain grounded throughout the experiment. (b) Simplified diagram of the Hall bar mesa and measurement setup. The mesa is outlined and top gates are shaded grey. Differential resistance measurements are indicated by labels connecting pairs of ohmic contacts.

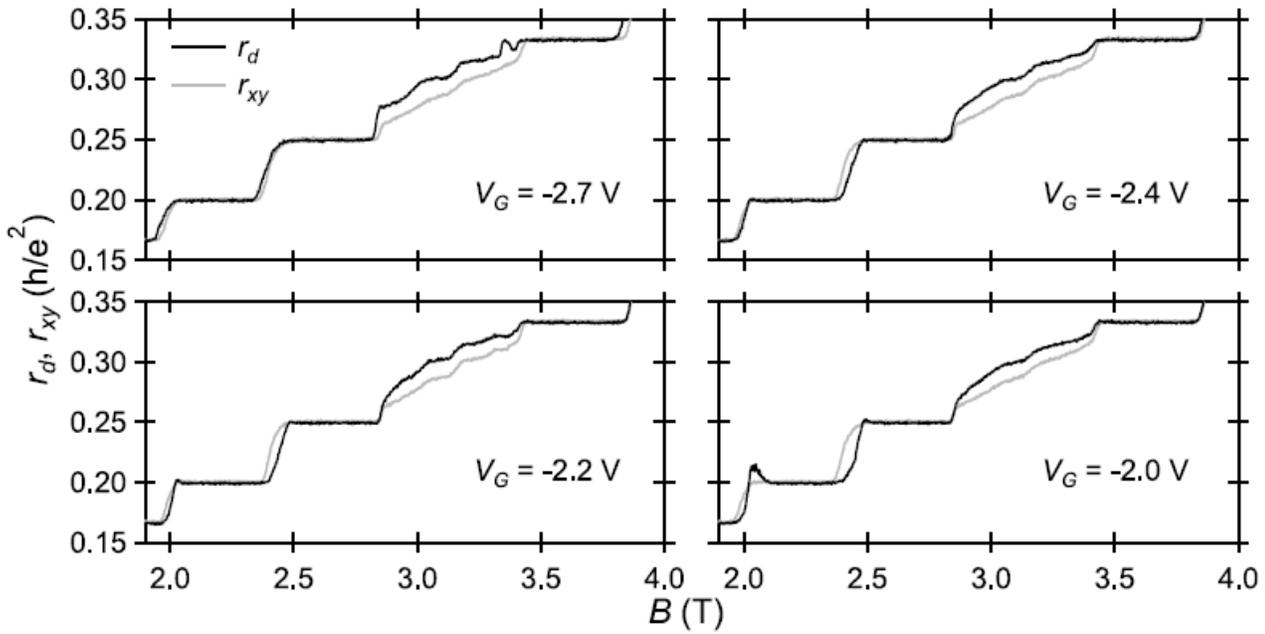

FIG. 2. Results of annealing and varying gate voltage thereafter. Measurements of $r_d$ and $r_{xy}$ after annealing at -2.7 V on the QPC gates. The slopes of $r_d$ (black) and $r_{xy}$ (grey) provide a measure of the electron density in the region of the QPC and far way in the Hall bar, respectively. The densities are closely matched at -2.7 V and matched to within 1% between -2.4 and -2.0 V.

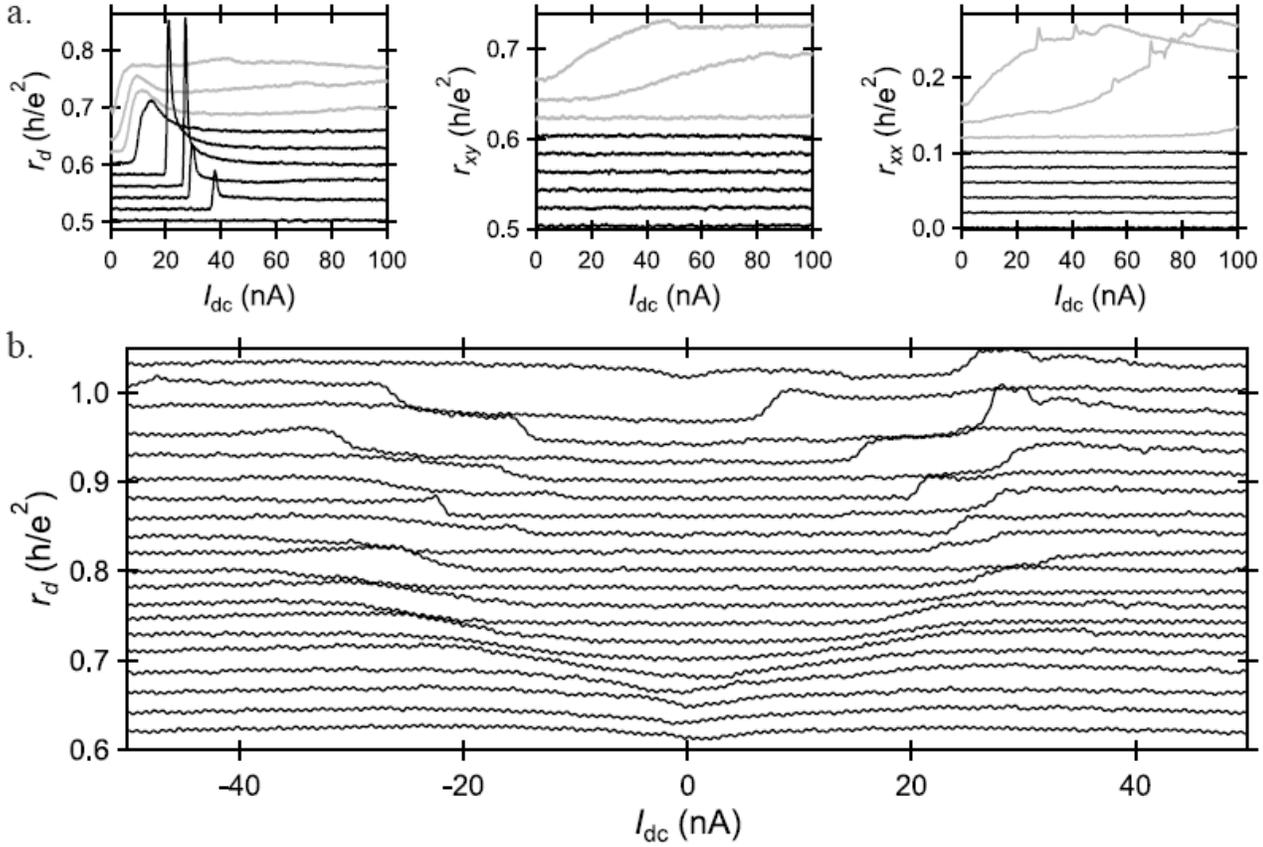

FIG. 3. Example of breakdown in the QPC at nonzero critical currents.

(a) Simultaneous measurements of $r_d$, $r_{xy}$, and $r_{xx}$ near $\nu = 2$. All curves are measured with increasing dc bias. Consecutive scans, from bottom to top, are separated by 80 mT (5420 to 6060 mT) and offset by 0.02 $h/e^2$ vertically. Black curves are those for which $r_{xy}$ remains quantized and $r_{xx}$ is zero within the noise over the entire range of dc bias. Grey curves are those for which $r_{xy}$ and/or $r_{xx}$ deviate from these values. Breakdown of the quantum Hall states in the QPC is characterized by a sharp increase in $r_d$ at a critical current. (b) Breakdown of the $\nu = 5/3$ plateau. Consecutive scans, from bottom to top, are separated by 20 mT (6300 to 6680 mT, covering slightly more than the 5/3 plateau) and offset by 0.02 $h/e^2$ vertically. The dc bias is swept in alternating directions for consecutive scans, with increasing $I_{dc}$ for the bottom scan.

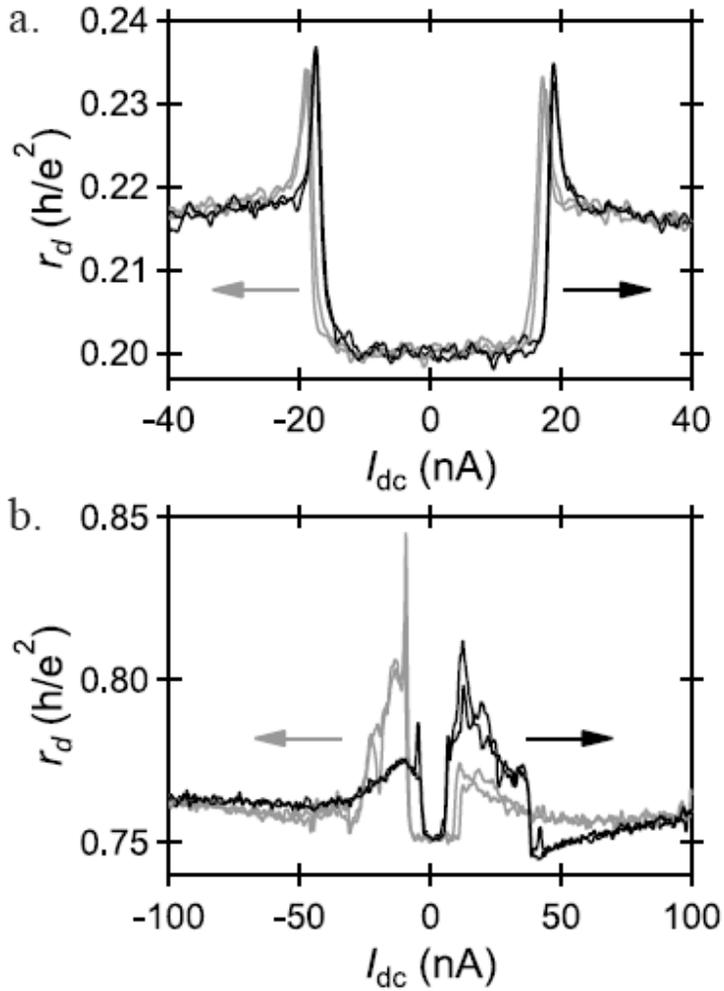

FIG 4. Examples of two different types of hysteresis.

(a) Hysteresis in which $I_c$ increases with increasing $I_{dc}$ and decreases with decreasing $I_{dc}$. (b) A different type of hysteresis in which $|I_c|$ is different for the two sweep directions. For both graphs, the curves are measured by continuously sweeping $I_{dc}$ in alternating directions and incrementing gate voltage by 4 mV between sweeps; arrows indicate the sweep direction of the corresponding curves. Curves in (a) are taken at 2.26 T ($\nu = 5$) with gate voltages -2.236 to -2.224 V. $I_{dc}$ is swept between ±100 nA but only displayed over a limited range to emphasize the hysteresis. Sweeps in (b) are taken at 8.15 T ($\nu = 4/3$) with gate voltages -2.280 to -2.268 V.

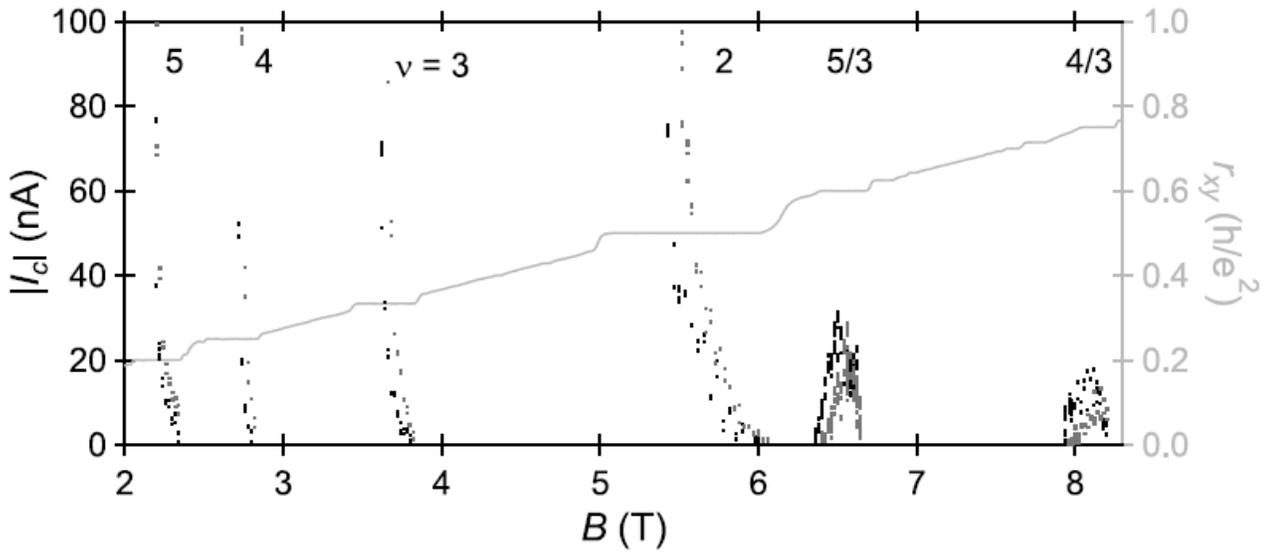

FIG. 5. Critical current as a function of magnetic field for various filling fractions.

Absolute value of the critical current plotted against magnetic field over the ν = 4/3, 5/3, 2, 3, 4, and 5 quantum Hall plateaus. Black dots are measured with a -2.4 V gate voltage and grey dots with -2.2 V. The light grey curve (right axis) shows a $r_{xy}$ Hall trace taken separately for reference. As explained in the text, breakdown on the low magnetic field side of the integer plateaus is qualitatively different and the critical current in these regions is not shown.

|   | $\alpha$ | |
| :---: | :---: | :---: |
| $\nu$ | $V_G = -2.4$ V | $V_G = -2.2$ V |
| 2 | 1.6 ± 0.2 | 2.1 ± 0.1 |
| 3 | 2.4 ± 0.2 | 1.2 ± 0.2 |
| 4 | 1.7 ± 0.4 | 1.4 ± 0.2 |
| 5 | 1.4 ± 0.2 | 1.2 ± 0.2 |

TABLE I. Exponents extracted by power law fits using Eq. (1) at integer filling factors.

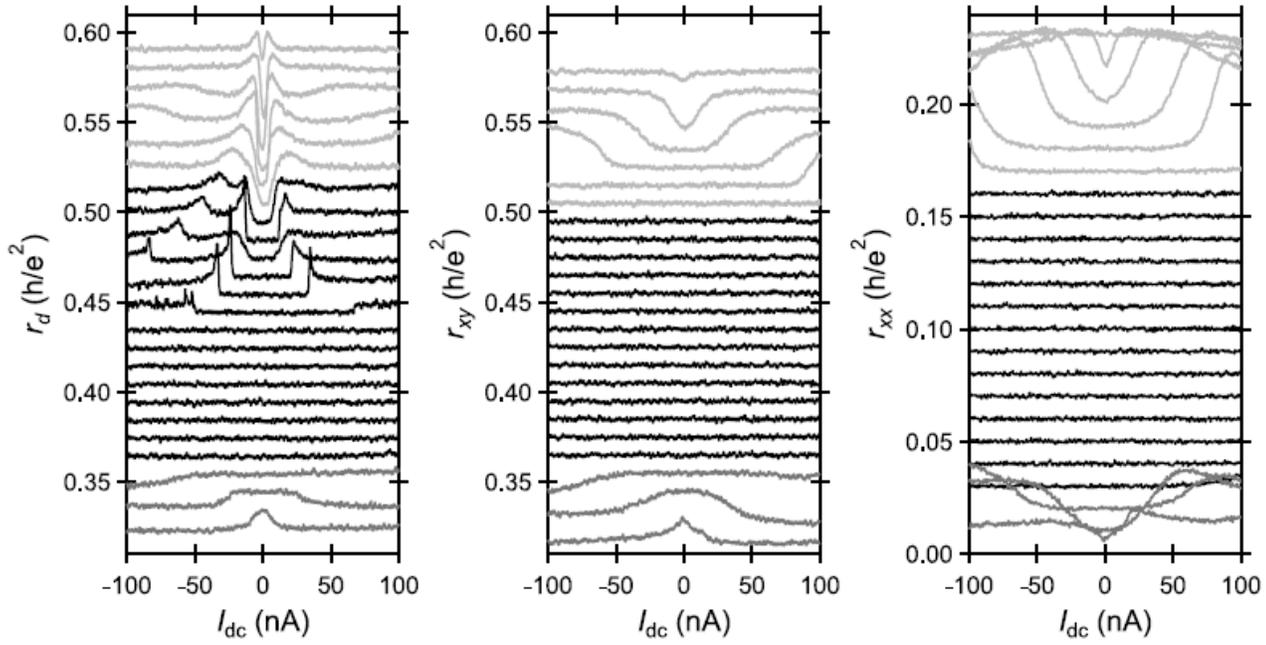

FIG. 6. Example of breakdown at an integer plateau.

Differential resistances $r_d$, $r_{xy}$, and $r_{xx}$ measured simultaneously as a function of dc bias around the $\nu = 3$ plateau. Consecutive scans, from bottom to top, are separated by 20 mT (3400 to 3840 mT) and offset by 0.01 $h/e^2$ vertically. The dc bias is swept in alternating directions for consecutive scans, with decreasing $I_{dc}$ for the bottom scan. Black curves are those for which $r_{xy}$ remains quantized and $r_{xx}$ is zero within the noise over the entire range of dc bias. Light grey curves are those for which $r_{xy}$ and/or $r_{xx}$ deviate from these values on the high magnetic field side of the plateau. The corresponding measurements of $r_d$ exhibit breakdown at much smaller critical currents. In contrast, on the low field side of the plateau all three resistances exhibit deviations from ideal behavior at similar values of dc bias and magnetic field (dark grey curves).

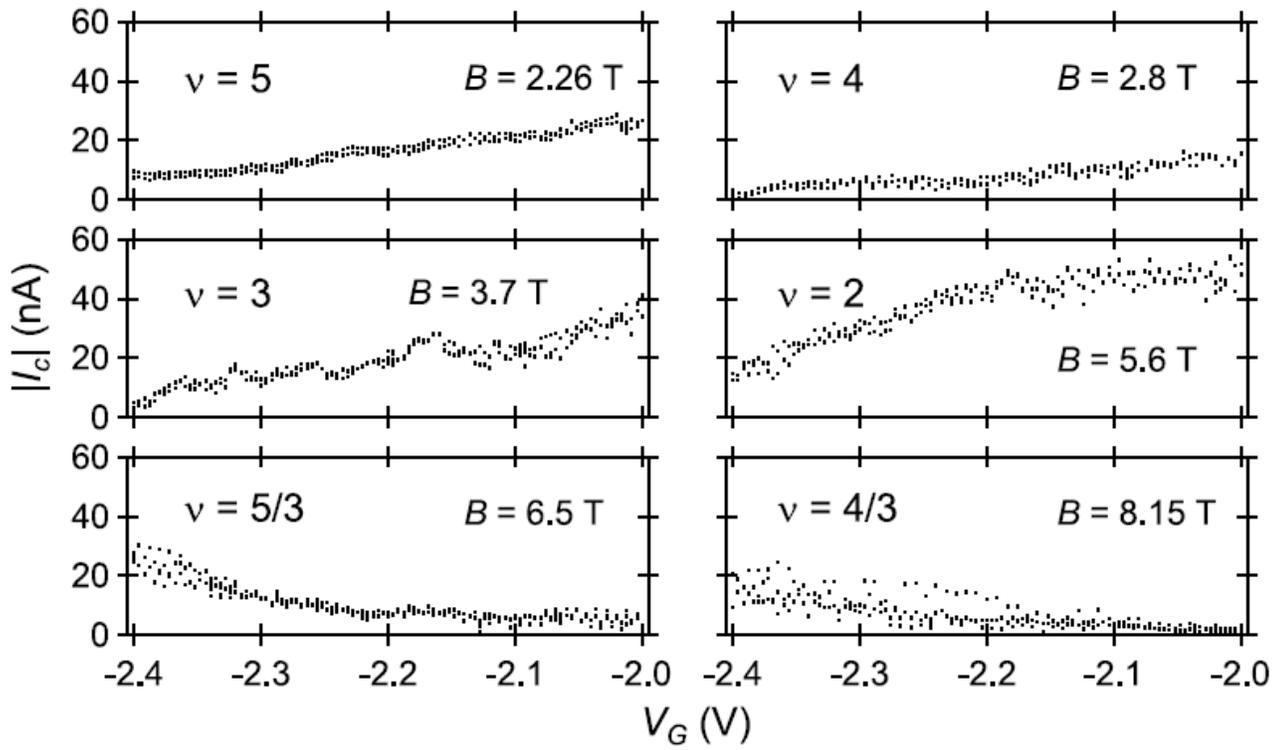

FIG. 7. Critical current as function of gate voltage for various filling factors. Absolute value of the critical current plotted against gate voltage for $\nu$ = 4/3, 5/3, 2, 3, 4, and 5. Each curve is measured at the indicated magnetic field. Integer filling factors exhibit critical currents increasing with more positive gate voltage while fractional filling factors exhibit decreasing critical currents.

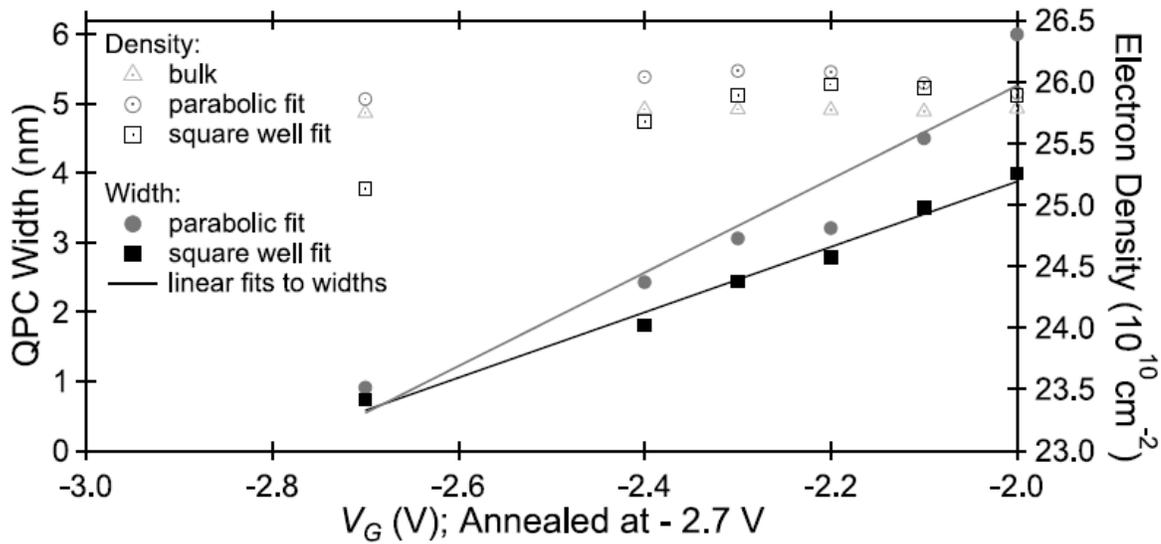

FIG. 8. QPC width and electron density as a function of gate voltage.

QPC width (left axis, solid symbols) as a function of gate voltage, estimated using two different models explained in the text. Lines are linear fits to each set of estimates. Electron density (right axis, open symbols) in the region of the QPC and in the Hall bar far from the QPC. Measurements of the electron density in the QPC are taken from the same fits which give the width estimates.

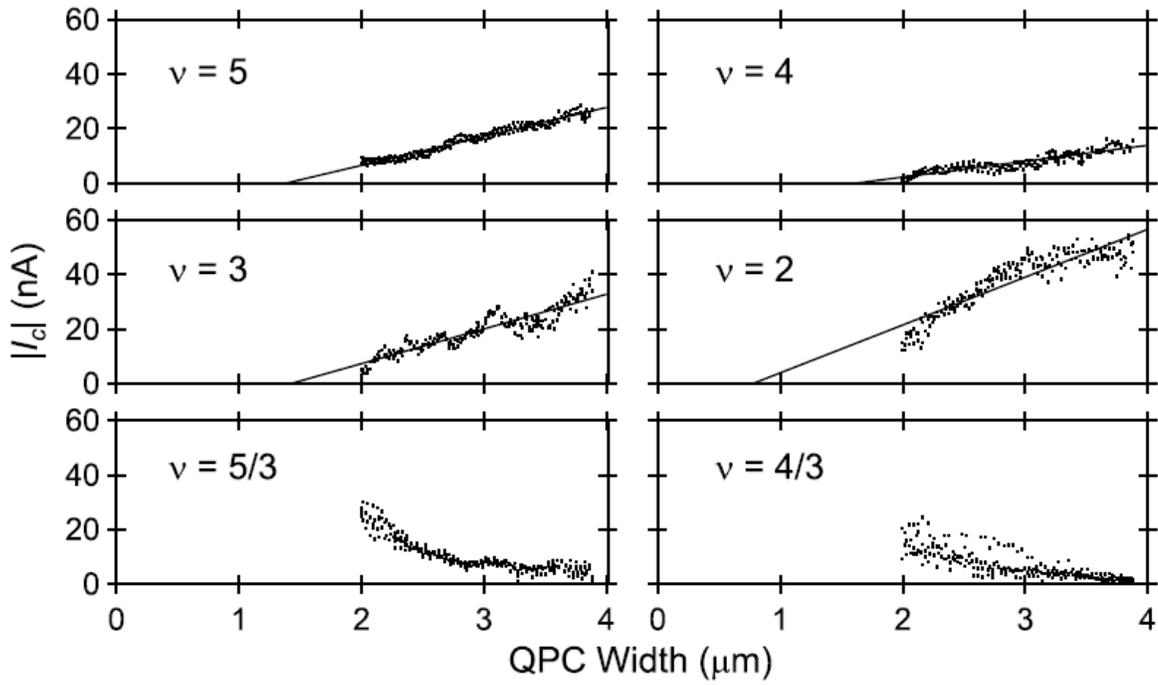

FIG. 9. Critical current as function of QPC width at various filling factors.

Results of Fig. 7 plotted against QPC width at the labeled filling factors. Widths are estimated as a function of gate voltage using the square well potential model (see the main text and Fig. 8). Results for the parabolic potential would increase all widths by ~30%. Lines are linear fits at each integer filling factor.